\documentclass[11pt,twoside]{article}
\usepackage{asp2004}
\usepackage{psfig}
\usepackage{epsf}
\usepackage{graphics}
\usepackage{lscape}
\def\edcomment#1{\iffalse\marginpar{\raggedright\sl#1\/}\else\relax\fi}
\reversemarginpar

\begin{document}
\title{Evolution of Magnetic Fields in Stars Across the Upper Main Sequence: 
Results from Recent Measurements with FORS\,1 at the VLT}
\author{S. Hubrig}
\affil{European Southern Observatory, Casilla 19001, Santiago 19, Chile}
\author{P. North}
\affil{Laboratoire d'Astrophysique de l'Ecole Polytechnique F\'ed\'erale
de Lausanne, Observatoire,
CH-1290~Chavannes-des-Bois, Switzerland}
\author{T. Szeifert}
\affil{European Southern Observatory, Casilla 19001, Santiago 19, Chile}

\begin{abstract}

We rediscuss the evolutionary state of upper main sequence magnetic stars
using a sample of Ap and Bp stars with accurate Hipparcos parallaxes
and definitely determined longitudinal magnetic fields. FORS\,1 at the VLT in 
spectropolarimetric mode has been used to carry out a systematic search for 
magnetic fields in chemically peculiar stars whose 
magnetic field has never been studied before.
We confirm our previous results obtained from the study of Ap and Bp stars
with accurate measurements of the mean magnetic field modulus and mean
quadratic magnetic fields that the distribution of the magnetic stars
of mass below $3M_\odot$ differs significantly from 
that of normal stars in the same temperature range. Normal A stars occupy
the whole width of the main sequence, without a gap, whereas magnetic stars 
are concentrated towards the centre of the main-sequence band. 
We show that, in contrast,  higher mass magnetic Bp stars may well occupy the 
whole main-sequence width.
\end{abstract}
\thispagestyle{plain}

\section{Introduction}
To properly understand the physics of 
upper main sequence stars it is particularly important to identify the origin 
of their magnetic fields. Two main streams of thought have been followed: one 
according to which the stars have acquired their field at the time of their
formation or early in their evolution (what is currently observed is then a 
fossil field), and the other according to which the field is  
generated and maintained by a contemporary dynamo.
Recently, we found that magnetic fields appear only in stars
of mass below $3M_\odot$ if they have 
already completed at least approximately 30\% of their main-sequence 
lifetime \citep{hu00}.
The absence of stars with 
strong magnetic fields close to the ZAMS might be seen as an argument against 
the fossil field theories. Yet, the whole sample under study            
contained only 33 magnetic stars
with accurate measurements of the mean magnetic field modulus or
mean quadratic magnetic fields from spectra taken in unpolarized light. 
For these stars the mean magnetic field modulus which is the average over the stellar 
disk of the modulus of the magnetic vector has been derived through the measurement of
the wavelength separation of resolved magnetically split components of spectral lines.
The mean quadratic field has been diagnosed 
from the consideration of the differential magnetic broadening of spectral lines. 
Our study suffered from a selection effect: our sample contained a 
high fraction (about 2/3) of stars with rotation periods longer than 10 days,
while the majority of the periods of magnetic stars fall between 2 and 4 days.  

The goal of our program currently scheduled on the VLT with 
FORS\,1 in spectropolarimetric mode is to
carry out a systematic search for longitudinal magnetic fields in 
about 100 upper main sequence chemically peculiar stars with good Hipparcos parallaxes 
in a wider range of masses whose 
magnetic field has been never or only poorly studied before, and with a distribution 
of rotation periods more representative of that of all Ap and Bp stars. 
The mean longitudinal magnetic field is the average over the stellar disk of the 
component of the magnetic vector along the line of sight and is derived from 
measurements of wavelength shifts between right and left circular polarization.
Here we present the first results of our study of the 
evolution of the magnetic field across the main sequence obtained
from the knowledge of the longitudinal 
magnetic fields and the accurate position of these stars in the H-R diagram.
To better constrain our results on the origin of the magnetic field in Ap and
Bp stars we enlarged
our data sample by including in this study also the data for
stars with accurate Hipparcos parallaxes and longitudinal fields reliably
measured in previous studies by different authors. The measurements of these additional
stars are compiled in the paper of \citet{by2003}.

\section{Basic data}
The General Catalogue of Ap and Am stars \citep{re96} includes
2875 Ap stars showing abnormal enhancement of one or several elements in their
atmospheres. Although Hipparcos parallaxes have been measured for about 940 Ap 
and Bp stars, only 371 of them have been measured at a low parallax error 
of $\sigma(\pi)/\pi<0.2$.
Currently 149 Ap stars have reliably measured longitudinal fields 
ranging from hundreds of Gauss to dozens of kG \citep{rk01}.
But only for 62 stars with measured magnetic fields the parallax error is
less than 20\%.
Half of these stars have in addition accurate measurements of the mean 
magnetic field modulus and mean
quadratic magnetic field and have been used in our previous study of the 
evolutionary state of magnetic stars. The mean field modulus and mean
quadratic magnetic field are, by definition,
much less aspect-dependent than the longitudinal field and, thus, they 
characterize the intrinsic stellar magnetic field much better. The observations 
also show that their variations are most often of low amplitude.
However, longitudinal field measurements represent the standard method of 
searching for magnetic fields in different types of stars, and 
all models of the geometry and detailed structure of the magnetic 
fields of these stars have been constrained using longitudinal field
measurements.
The distribution of the other half of the stars for which longitudinal fields
in the H-R diagram have been reliably measured is
shown in Fig.\,1 (left). 


\begin{figure}
\plottwo{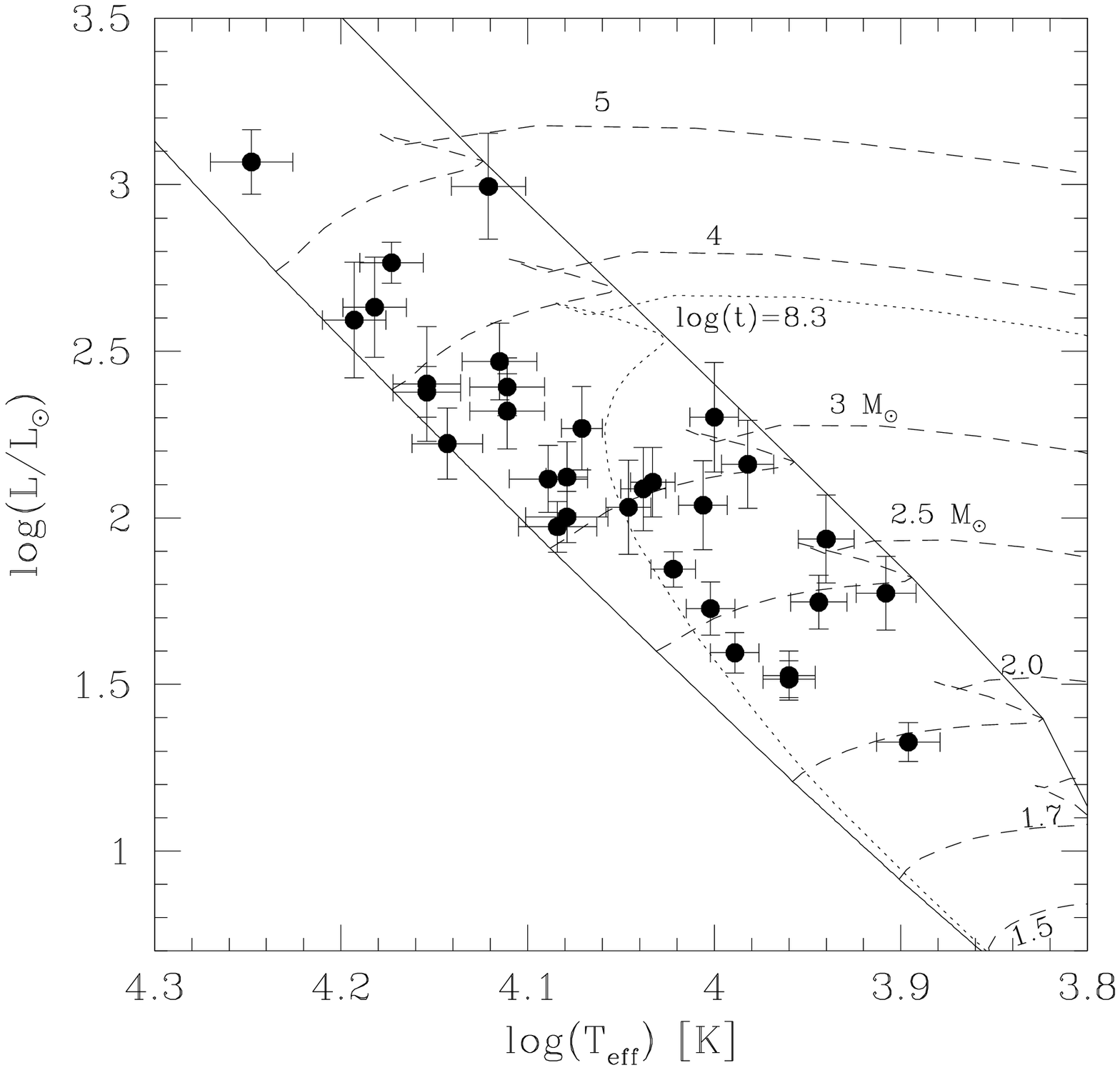}{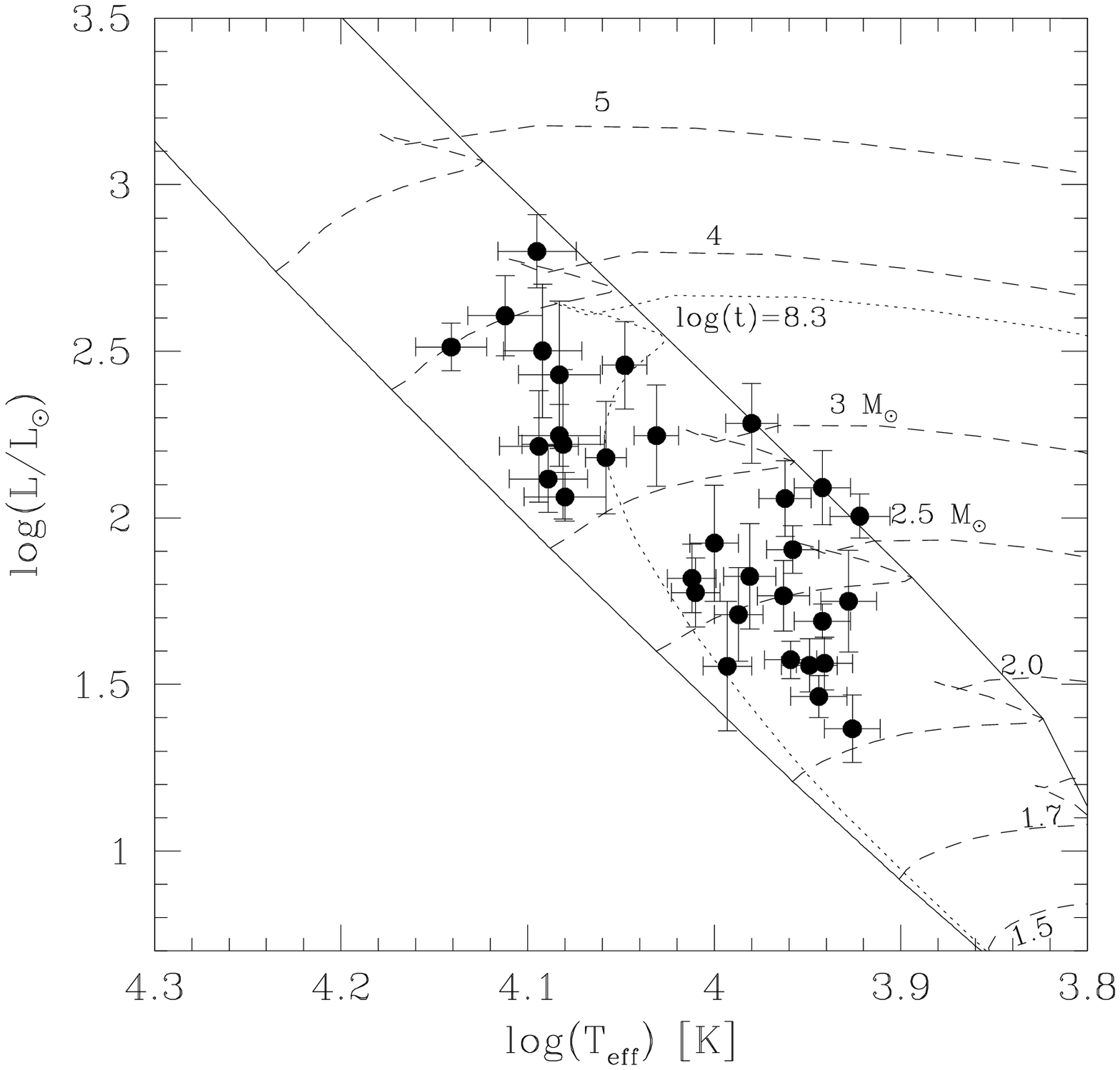}
\caption{
{\bf Left:} The H-R diagram for the sample of magnetic stars with already known
mean longitudinal fields.
{\bf Right:} The H-R diagram for the sample of magnetic stars with 
mean longitudinal fields measured with FORS\,1.
}
\end {figure}

Using FORS\,1 in service mode during the year 2003, new longitudinal field
determinations have been obtained for approximately one third of our
star sample. FORS\,1 is a multi-mode instrument which is equipped with 
polarization analyzing optics comprising super-achromatic half-wave and 
quarter-wave phase 
retarder plates, and a Wollaston prism with a beam divergence of 22$\arcsec$ 
in standard resolution mode. For all stars we used the GRISM\,600B in the 
wavelength range 3480--5890\,\AA{} to cover all hydrogen Balmer lines from 
H$\beta$ to the Balmer jump. The determination of the mean longitudinal fields
using FORS\,1 is described in detail in \citet{hu04}.
The distribution of the magnetic stars with the longitudinal fields measured with
FORS\,1 in the H-R diagram is shown in Fig.\,1 (right). 


The effective temperatures of the stars of both samples have been determined
from photometric data in the Geneva system (e.g., \cite{k97}) or, when no
Geneva photometry was available, in the
Str\"omgren system, applying the calibration of \citet{mn85}.
The luminosities have been obtained by taking into account the bolometric
corrections measured by \citet{La84}.
In case of binary systems, a correction for the duplicity has been applied to their
magnitudes (i.e. $0.75$~mag for SB2 and $0.3$~mag for SB1 systems).
The Lutz--Kelker correction \citep{lk73} has been brought, taking
into account a stellar density varying as a function of the galactic latitude.
More details of the determination of the fundamental parameters can be found
in our previous paper \citep{hu00}.

\section{Analysis and results}

It is obvious from Fig.\,1 that magnetic stars of mass below $3M_\odot$ are only 
rarely found close to the zero-age main sequence, supporting the view that magnetic
Ap stars are observed only in a restricted range of evolutionary states.
The majority of the rotational periods of the studied stars fall between 2 and 4 days, 
and there is no indication that the distribution of these stars in the H-R diagram is
different than that of very slowly rotating magnetic stars. 

By contrast, the stars of higher mass seem to fill the whole width of the 
main-sequence band. 
However, we should note that for Bp stars the effective temperatures 
derived from photometry are not in good agreement with the spectral 
classification \citep{hu00}. 
Because of the extremely anomalous energy distribution and large variations of 
their spectra, the calibration of the photometric temperature indicators are
frequently questioned. The goal of our future work is to try to resolve these
inconsistencies by detailed spectroscopic studies of these stars. While  
a few double--lined spectroscopic binary systems 
containing an Ap star of mass  below $3M_\odot$ are currently known, the rate of binaries
is much smaller among magnetic Bp stars \citep{g85}, and 
only one double--lined eclipsing binary with a Bp component, namely AO Vel, is known 
to date. The evolutionary state of this star is currently under study by P. North.

Because of the strong dependence of the longitudinal field on the rotational 
aspect, its usefulness to characterise actual field strength distributions is 
limited, but this can be overcome, at least in part, by repeated observations to 
sample various rotation phases, hence various aspects of the field. 
Three observations per star should be the strict minimum to 
be able to apply in a meaningful way the kind of statistics 
we use to confirm the detection of a field from 
longitudinal field measurements based on the rms longitudinal field 
strength computed from all the measurements (see eqn.\ 1) and a reduced
chi-square statistics.

\begin{equation}
\left< B_l^2 \right>^{1/2} = \left( \frac{1}{n_1} \sum^{n_1}_{i=1} B^2_{li} \right)^{1/2}
\end{equation}
%


In Fig.\,2 (left) the rms longitudinal field strength is plotted against the 
age (expressed as a fraction of their total main-sequence life time) of stars 
with magnetic fields already known from previous studies. 
Only one or two measurements are currently available for the stars observed
with FORS\,1. Some of them show mean longitudinal fields below the
3$\sigma$ level 
and it still must be established if all of them have detectable magnetic 
fields.
The measured longitudinal fields in the stars observed with FORS\,1
as a function of the completed fraction of main-sequence life are presented in Fig.\,2 (right).
\begin{figure}
\plottwo{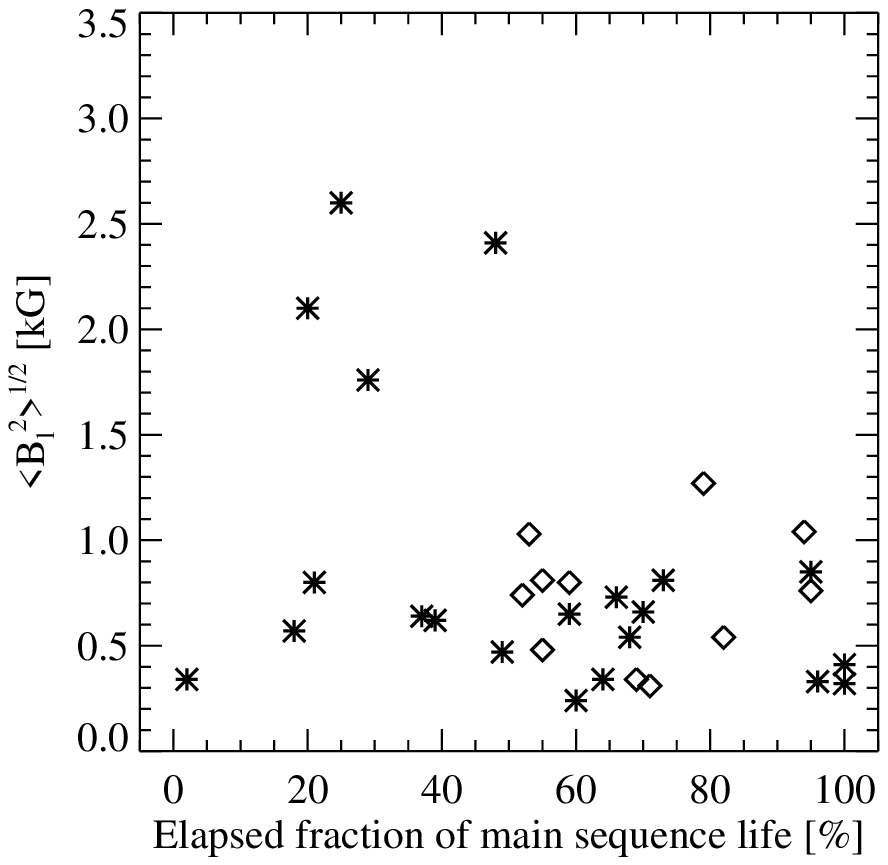}{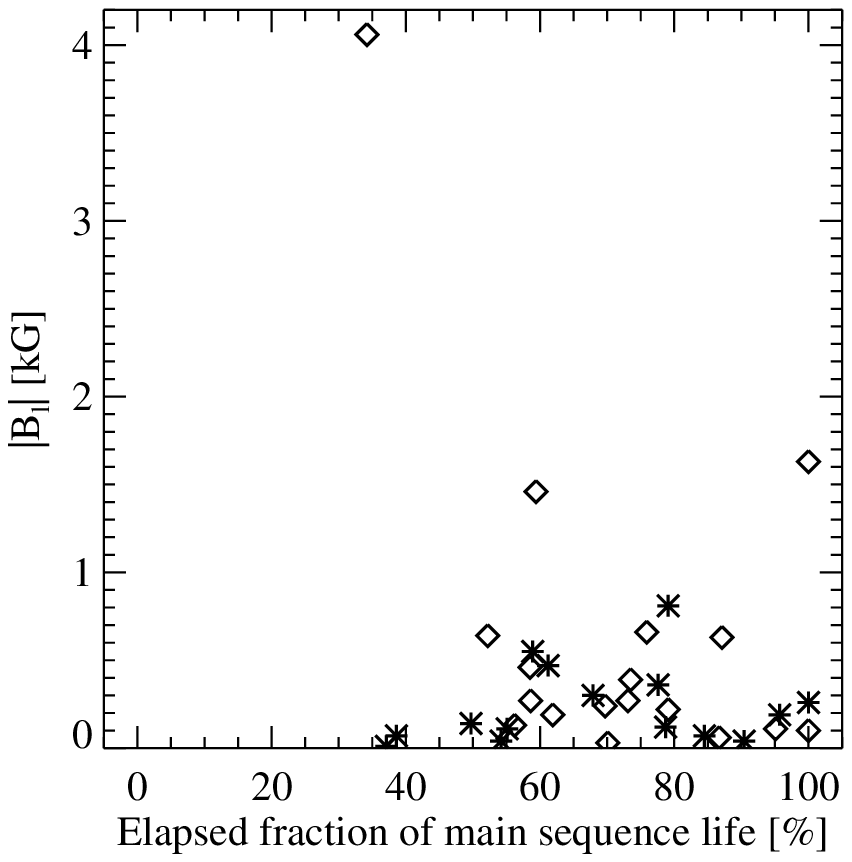}
\caption{{\bf Left:} The rms longitudinal field strength as a function of the 
completed fraction of main-sequence life for stars with reliably measured longitudinal 
fields. {\bf Right:} The mean longitudinal field versus age for the stars observed with
FORS\,1. Asterisks correspond to the stars of mass above $3M_\odot$ whereas diamonds
distinguish stars of lower mass.}
\end {figure}

It is clearly seen from Fig.\,2 that magnetic fields become observable in the lower
mass stars only after they have completed a significant fraction of their life on
the main sequence, more than 30\%.
Our results also show that stronger magnetic fields tend to be found in hotter, 
younger (in terms of the elapsed fraction of main-sequence life) and more massive 
stars. \citet{hu00} have 
already reported about the existence of such a trend in their study of the 
evolutionary state of magnetic Ap stars. 

Certainly, further systematic studies of magnetic fields in Ap and Bp stars should be 
conducted with a view to derive unambiguous results about the origin of the magnetic 
fields of the Ap and Bp stars. Only a few double--lined spectroscopic binary systems 
containing a magnetic Ap star are currently known,  and in all the studied systems
the Ap components have already completed a significant fraction of the main-sequence life.
As far as the membership of Ap stars in 
distant open clusters is concerned,
we should keep in mind that such studies are mostly based upon photometry
and upon radial velocity determinations. However, criteria for assessing 
cluster membership based on photometry cannot be  applied to peculiar stars
straight away, in
which strong backwarming effects lead to an anomalous energy distribution, thus
affecting the position of the star in colour--magnitude diagrams.
12 stars in our sample are known members
of nearby open clusters of different ages and have very accurate Hipparcos parallaxes.
They are very promising candidates for our study and the measurements of their magnetic
fields will allow us to put more 
stringent constraints on the origin of the magnetic fields. The study of these stars
is currently under way.

\vfill\pagebreak
\pagebreak


\begin{thebibliography}{}
\bibitem[Bychkov et al.(2003)]{by2003}
Bychkov, V. D., Bychkova, L. V., \& Madej, J. 2003, A\&A 407, 631
\bibitem[Gerbaldi et al.(1985)]{g85}
Gerbaldi, M., Floquet, M., \& Hauck, B. 1985, A\&A 146, 341
\bibitem[Hubrig, North \& Mathys(2000)]{hu00}
Hubrig, S., North, P., \& Mathys, G. 2000, ApJ 539, 352
\bibitem[Hubrig et al.(2004)]{hu04}
Hubrig, S.\ et al.\ 2004, A\&A 415, 661
\bibitem[K\"unzli et al.(1997)]{k97}
K\"unzli, M.\ et al.\ 1997, A\&AS 122, 51
\bibitem[Lanz(1984)]{La84} 
Lanz, T.\ 1984, A\&A 139, 161
\bibitem[Lutz \& Kelker(1973)]{lk73}
Lutz, T.E. \& Kelker, D.H. 1973, PASP 85, 573
\bibitem[Moon \& Dworetsky(1985)]{mn85}
Moon, T.T. \& Dworetsky, M.M. 1985, MNRAS 217, 305
\bibitem[Renson et al.(1984)]{re96}
Renson, P., Gerbaldi, M., \& Catalano, F. A. 1996, VizieR On-line Data Catalog: 
III/162A. 
\bibitem[Romanyuk \& Kudryavtsev(2001)]{rk01}
Romanyuk,  I. I. \& Kudryavtsev, D.O. 2001, in ASP Conf.\ Ser.\ Vol.\ 248, 299
\end{thebibliography}
\end{document}